\documentclass[useAMS,usenatbib]{mn2e}

\newcommand{\gtsim}{\protect\raisebox{-0.5ex}{$\:\stackrel{\textstyle >}
        {\sim}\:$}}

\usepackage{longtable}    
\usepackage{graphicx}
\usepackage{epsfig}
\usepackage{rotating}
\usepackage{lscape}
\usepackage{latexsym}
\usepackage{psfig}

\title[Abundances of low-mass stars in nearby young associations]{Elemental abundances of low-mass stars in nearby young 
associations: AB Doradus, Carina Near, and Ursa Major\thanks{Based on 
observations performed with European Southern Observatory telescopes (program IDs: 70.D-0081(A), 082.A-9007(A), 083.A-9011(B), 
084.A-9011(B)).}}
\author[K. Biazzo et al.]{K. Biazzo$^{1}$\footnotemark[0]\thanks{E-mail: katia.biazzo@oacn.inaf.it}, V. D'Orazi$^{2,4}$, 
S. Desidera$^{3}$, E. Covino$^{1}$, J. M. Alcal\'a$^{1}$, M. Zusi$^{1}$\\
$^{1}$INAF - Capodimonte Astronomical Observatory, via Moiariello, 16, 80131 Naples, Italy\\
$^{2}$Department of Physics \& Astronomy, Macquarie University, Balaclava Rd., North Ryde, Sydney, NSW 2109, Australia\\
$^{3}$INAF - Padova Astronomical Observatory, vicolo dell'Osservatorio, 5, 35122 Padova, Italy\\
$^{4}$Monash Centre for Astrophysics, School of Mathematical Sciences, Building 28, Monash University, VIC 3800, Australia}
\begin{document}

\date{Accepted 2012 Month XX. Received 2012 Month XX.}

\pagerange{\pageref{firstpage}--\pageref{lastpage}} \pubyear{2012}

\maketitle

\label{firstpage}

\begin{abstract}
{We present stellar parameters and abundances of 11 elements (Li, Na, Mg, Al, Si, Ca, Ti, Cr, Fe, Ni, and Zn) of 13 F6-K2 
main-sequence stars in the young groups AB~Doradus, Carina Near, and Ursa Major. The exoplanet-host star $\iota$\,Horologii is 
also analysed. 

The three young associations have lithium abundance consistent with their age. All other elements show solar abundances. 
The three groups are characterised by a small scatter in all abundances, 
with mean [Fe/H] values of 0.10 ($\sigma=0.03$), 0.08 ($\sigma=0.05$), and 0.01 ($\sigma=0.03$) dex for AB~Doradus, 
Carina Near, and Ursa Major, respectively. The distribution of elemental abundances appears congruent with 
the chemical pattern of the Galactic thin disc in the solar vicinity, as found for other young groups.
This means that the metallicity distribution of nearby young stars, targets of direct-imaging planet-search surveys, 
is different from that of old, field solar-type stars, i.e. the typical targets of radial velocity surveys.

The young planet-host star $\iota$\,Horologii shows a lithium abundance lower than that found for the young association members.
It is found to have a slightly super-solar iron abundance ([Fe/H]=0.16$\pm$0.09), while all [X/Fe] ratios are similar to the 
solar values. Its elemental abundances are close to those of the Hyades cluster derived from the literature, which seems to reinforce 
the idea of a possible common origin with the primordial cluster. 
}
\end{abstract}


\section{Introduction}
During the last twenty years, a dozen of young ($<$500 Myr) nearby ($<$200 pc) associations (or co-moving stellar groups) 
have been identified (see, e.g., \citealt{montesetal2001}, \citealt{zuckermanetal2004}, \citealt{torresetal2008}). 
Although numerous kinematical studies have confirmed their existence, their origin and evolution remain still unclear 
(see \citealt{liuetal2012}, and references therein). 
Representing valuable laboratories to investigate the recent star formation in the solar vicinity, the measurement and study of their 
chemical composition are important to put constraints on their origin and evolutionary history, but also for the exo-planetary research. 
On one side, elemental abundances of $\alpha$-elements (but also iron-peak elements) in young associations can provide evidence 
of recent local enrichment; on the other side, since planets are assumed to form from circumstellar discs during the pre-main sequence 
phase, obvious questions arise on what the metallicity of young solar analogs and what fraction of them (if any) is metal-rich 
(see, e.g., \citealt{biazzoetal2011a}, and references therein). 
Yet, so far, only a few studies have been focused on the determination of elemental abundances in such stellar groups 
(see Section \ref{sec:obs}).

Many studies have shown that giant gaseous planets are preferentially found around main-sequence solar-type stars more metal-rich than 
the Sun (e.g., \citealt{johnsonetal2010}, and references therein). In particular, the frequency of giant planets around stars of twice 
the solar metallicity is around 30\%, in contrast to the $\sim 3$\% for stars with solar or sub-solar iron content 
(e.g., \citealt{fischervalenti2005, sousaetal2011}). Such a trend seems to have a primordial/basic origin (\citealt{gillietal2006}), 
but presents several caveats. First, giant stars hosting planets do not appear on average more metal-rich than stars without planets 
(\citealt{pasquinietal2007}), and this could hint that stellar mass strongly influences the planet formation process (\citealt{santosetal2012}). 
Second, the trend is no longer valid for iron abundances ranging from [Fe/H]=$-0.7$ to [Fe/H]=$-0.3$ dex (\citealt{haywood2009}). 
Third, the nature of such a trend in the early stages of planet formation is still unknown. Regarding the latter point, the 
dispersal efficiency of circumstellar (or proto-planetary) discs, the planet birthplace, is predicted to depend on metallicity 
(\citealt{ercolanoclarke2010}). In a recent study, \cite{yasuietal2010} have found that the disc fraction in significantly low-metallicity 
clusters ([O/H]$\sim -0.7$) declines much faster (in $<1$~Myr) than observed in solar-metallicity clusters (i.e. in $\sim 5-7$ Myr).
They suggest that, as the shorter disc lifetime reduces the time available for planet formation, 
this could be one of the reasons for the strong planet-metallicity connection.

In this paper, we investigate the abundances of 11 elements (lithium, iron-peak, $\alpha$, and other 
odd-/even- Z elements) in 13 F6--K2 main-sequence stars belonging to the young associations AB~Doradus, Carina Near, and Ursa Major. 
The case of $\iota$\,Horologii (a young exoplanet-host star) is also investigated. Some of these associations were already studied 
in terms of some elemental abundances by several authors (e.g., \citealt{desideraetal2006b, paulsonyelda2006, vianaetal2009, 
ammler-guenther2009}), but no effort has been done to widely characterise their chemical content. Recently, in a companion 
paper, we have reported the $s$-process element (yttrium, zirconium, lanthanum, cerium, and barium) abundance determination 
of the same targets in our sample (with the only exception of HIP\,36414), with the aim to investigate possible over-abundances 
(\citealt{dorazietal2012}). We have found that while Y, Zr, La, and Ce exhibit solar ratios, Ba is over-abundance by $\sim$0.2 dex; 
we have hence exploited effects related to the stratification in temperature of model atmospheres, NLTE corrections, and chromospheric-related 
effects as possible explanations for this scenario. Thus, the study of \cite{dorazietal2012} and the present complementary work represent 
the first efforts done to systematically derive many elemental abundances in young associations.

A brief overview of previous investigations in the selected young associations is given in the following of this Section. 
Section~\ref{sec:obs} presents the selection of the stellar sample and observations. Abundance analysis techniques are described in 
Section~\ref{sec:abundance}, while the results are presented in Section~\ref{sec:results} and discussed in Section~\ref{sec:discussion}. 
Summary and conclusions are given in Section \ref{sec:conclusions}.

\subsection{The AB Doradus group}
\label{sec:ab_dor}
The AB~Doradus (hereafter, AB~Dor) stellar group was first postulated by \cite{torresetal2003} in the SACY (Search for Associations 
Containing Young stars) project with the designation of AnA, and then identified by \cite{zuckermanetal2004} as the co-moving youthful 
($\sim 50$ Myr) group closest to Earth. They also claimed its nucleus is a clustering of a dozen F--M type members $\sim 20$ pc from 
Earth that includes the ultra-rapid rotator, active binary star AB~Dor. \cite{luhmanetal2005} argued that the AB~Dor association 
is a remnant of the large-scale star formation event that formed the Pleiades, and estimated an age of 75--150 Myr (this older age was 
also confirmed by \citealt{messinaetal2010}). The common origin of the AB~Dor and the Pleiades associations has been later reinforced 
by \cite{ortegaetal2007}. Recently, \cite{torresetal2008} presented the 89 high-probable members of AB~Dor, of which 29 are binaries, 
and derived a distance of $34\pm26$ pc and an age of 70 Myr. More recently, \cite{zuckermanetal2011} and \cite{schliederetal2012} 
found other likely members of the AB~Dor group, which include early-type stars, an M dwarf triple system, and three very cool 
objects.

\subsection{The Carina Near group}
The Carina Near association was identified by \cite{zuckermanetal2006} as a group of about 20 co-moving 200$\pm$50~Myr 
old stars, where all but three are plausible members of multiple stellar systems. The nucleus, at $\sim$30 pc from Earth, seems to be 
farther than the surrounding stream stars, and is located in the southern hemisphere and coincidentally quite close to the nucleus 
of the AB~Dor group, notwithstanding that the two groups have different ages and Galactic space motions 
(\citealt{zuckermanetal2004, zuckermanetal2006}). 

\subsection{The Ursa Major group}
The Ursa Major (hereafter, UMa) association in the Big Dipper constellation is located at a distance of $\sim$25 pc. It includes 
$\sim 50$ most probable members placed across almost the whole northern sky that move toward a common convergent point. 
The age estimates range widely from 200~Myr to 600~Myr (see \citealt{ammler-guenther2009}, and references therein).

\subsection{The exoplanet-host star $\iota$\,Horologii}
\label{sec:iota_Hor}
The young exoplanet-host star $\iota$\,Horologii (\citealt{kursteretal2000}) has been studied by many authors during the last decade, 
in particular for the implications on theories of stellar and planetary formation and possible relationship with metallicity. 
It belongs to the Hyades stream (\citealt{vauclairetal2008}), which is composed by field-like stars (85\%) and stars evaporated from 
the primordial Hyades cluster (15\%). Recent asteroseismic studies suggest that $\iota$\,Hor was formed within the primordial 
$\sim 600$\,Myr-old Hyades cluster and then evaporated toward its present location, 40\,pc away (see \citealt{vauclairetal2008}, 
and references therein). The same studies show that the metallicity, helium abundance, and age are similar to those of the Hyades 
cluster.

\setlength{\tabcolsep}{2.5pt}
\begin{table*}
\begin{center}
  \caption{Data from the literature.\label{tab:literature}}
  \begin{tabular}{@{}lcccclccrclr@{}}
  \hline
   Star & $\alpha$ & $\delta$ & $V$ & $B-V$ & SpT & $\pi$ & $V_{\rm rad}$ & $EW_{\rm Li}$ & $v\sin i$ & Note & 
   References$^a$\\
	& (hh:mm:ss) & ($\degr$:':'') & (mag) & (mag)& & (mas) & (km s$^{-1}$) & (m\AA) & (km s$^{-1}$) & \\
 \hline
\multicolumn{11}{c}{AB Doradus}\\
TYC\,9493-838-1  & 07:30:59.5 & $-$84:18:27.8 &9.96&0.86& G9 & 14.1 &	    24.2$\pm$0.8 & 300  &  3.0$\pm$0.1 & &  [4],[8] \\
HIP\,114530      & 23:11:52.1 & $-$45:08:10.6 &8.80&0.72& G8 & 20.3$\pm$1.1 &	  11.2$\pm$1.3 & 220  &  6.6$\pm$1.2 & Binary &  [2],[4],[6] \\
TYC\,5155-1500-1 & 19:59:24.2 & $-$04:32:06.2 &9.43&0.75& G5 & 10.5 &		   & 140  &  9.0$\pm$2.0 & &  [4],[8],[10] \\
HIP\,82688       & 16:54:08.1 & $-$04:20:24.7 &7.82&0.60& G0 & 21.4$\pm$0.9 &$-$16.91$\pm$2.20& 133  & 16.8$\pm$0.6 & &  [2],[6],[7],[9] \\
TYC\,5901-1109-1 & 05:06:27.7 & $-$15:49:30.4 &9.12&0.63& F8 & 12.3 &		   & 140  &  6.0$\pm$2.0 & &  [4],[8],[10] \\
 \hline  
\multicolumn{11}{c}{Carina Near}\\
HIP\,37923         & 07:46:17.0 & $-$59:48:34.1 &8.23&0.83& K0 & 33.5$\pm$3.5$^{*}$  & 17.0$\pm$1.0 &  76 & 3.2$\pm$1.2 & Wide Binary & [3],[5],[6] \\
HIP\,37918         & 07:46:14.8 & $-$59:48:50.7 &8.14&0.78& K0 & 25.7$\pm$1.7$^{*}$  & 17.0$\pm$1.0 & 110 & 6.3$\pm$1.2 & Wide Binary & [3],[5],[6] \\
HIP\,58241$^{***}$ & 11:56:43.8 & $-$32:16:02.7 &7.81&0.67& G3 & 34.3$\pm$6.4$^{**}$ &  6.7$\pm$0.3 & 110 & 9.0$\pm$1.2 & Wide Binary & [5],[11],[12] \\
HIP\,58240$^{***}$ & 11:56:42.3 & $-$32:16:05.4 &7.64&0.64& G3 & 28.6$\pm$6.4$^{**}$ &  6.0$\pm$0.4 & 111 & 5.2$\pm$1.2 & Wide Binary & [5],[11],[12] \\
HIP\,36414         & 07:29:31.4 & $-$38:07:21.6 &7.74&0.52& F7 & 19.0$\pm$0.5 & 28.0$\pm$2.0 &  80 &   & Single & [5],[6] \\
 \hline  
\multicolumn{11}{c}{Ursa Major}\\
HD\,38392 ($\gamma$~Lep~B) & 05:44:26.5 & $-$22:25:18.8 &6.15&0.94& K2 & 112.0 & $-$9.57$\pm$0.13 &      & 2.8$\pm$1.8 &	   & [1],[3],[6] \\
HIP\,27072 ($\gamma$~Lep~A) & 05:44:27.8 & $-$22:26:54.2 &3.60&0.47& F6 & 112.0 & $-$9.22$\pm$0.12 &      & 7.7$\pm$1.8 &	   & [1],[3],[6] \\
 \hline  
\multicolumn{11}{c}{Hyades stream? }\\
HIP\,12653 ($\iota$ Hor)& 02:42:33.5 & $-$50:48:01.1 &5.40&0.57& F8 &  58.3 & 16.7$\pm$0.2 &  40 & 6.2$\pm$1.2 &   & [4],[6] \\
\hline  
\end{tabular}
\end{center}
\small{$^a$\,[1]: \cite{montesetal2001}; [2]: \cite{zuckermanetal2004}; [3]: \cite{desideraetal2006a}; [4]: \cite{torresetal2006}; 
[5]: \cite{zuckermanetal2006}; [6]: \cite{vanleeuwen2007}; [7]: \cite{whiteetal2007}; [8]: \cite{torresetal2008}; 
[9]: \cite{guilloutetal2009}; [10]: \cite{daSilvaetal2009}; [11]: \cite{tokovinin2011}; [12]: \cite{andersonetal2012}.\\
$^*$, $^{**}$ {\it Hipparcos} parallaxes of visual binaries have often large errors. Physical association between the components 
is confirmed by common proper motions and radial velocities.
\\
$^{***}$ HIP\,58240B=HIP\,58241 have a possible close companion (\citealt{tokovinin2011}) that is too faint to contribute
significantly to the optical spectrum and affect our abundance analysis.
}
\end{table*}

\section{Sample selection, spectroscopic database, and data reduction}
\label{sec:obs}

In this work, we determine elemental abundances of 13 confirmed members of young moving groups, distributed as follows (see 
Table~\ref{tab:literature}):
\begin{itemize}
\item Five stars belong to AB\,Doradus. Two of them (namely, HIP\,114530 and TYC\,9493-838-1) were also analysed by \cite{vianaetal2009} 
within the SACY project, in terms of iron, silicon, and nickel abundances. Thus, they can be used as comparison targets.
\item Five stars in the Carina Near group. Three stars (namely, HIP\,36414, HIP\,37198, and HD\,37923), with radial velocities in the range 
$\sim 17-28$~km~s$^{-1}$, belong to the cluster nucleus, while HIP\,58240 and HIP\,58241 are probable ``stream'' members with 
a radial velocity of $\sim$6~km~s$^{-1}$. Recently, \cite{desideraetal2006b} estimated the iron abundance of HD\,37923 and HD\,37918.
\item Two stars in the Ursa Major group, namely $\gamma$~Lep~A (HIP\,27072) and $\gamma$~Lep~B (HD\,38392). Only their iron abundance has 
been recently measured by \cite{paulsonyelda2006} and \cite{ramirezetal2007}.
\item $\iota$ Horologii, for which the abundances of a few elements were derived in the recent past (see Table~\ref{tab:param_chem}).
\end{itemize}

The sample was selected according to the following criteria: 
\begin{itemize}
\item dwarf stars with spectral types from late-F to early-K. Later spectral types were excluded because at effective temperatures lower 
than $\sim$4500 K significant formation of molecules occurs and abundance determinations through line equivalent widths become unreliable;
\item stars with projected rotational velocity lower than 15 km s$^{-1}$, to avoid line blending due to rotational broadening;
\item no double-lined spectroscopic binary;
\item no close visual binary to avoid contamination in the spectrum.
\end{itemize}
 
For our analysis we exploited stellar spectra obtained with FEROS (\citealt{kauferetal1999}) at the ESO/MPG 2.2m telescope. 
Five spectra were acquired as part of the program aimed at the spectroscopic characterisation of targets for the 
SPHERE\footnote{SPHERE (Spectro-Polarimetric High-contrast Exoplanet REsearch) is a next generation instrument that will have as 
prime objective the discovery of extra-solar giant planets orbiting nearby stars by direct imaging.}@ESO GTO survey 
(\citealt{mouilletetal2010}); the spectra of HIP\,27072 ($\gamma$~Lep A) and HD\,38392 ($\gamma$~Lep\,B) were taken from 
\cite{desideraetal2006a}, while the remaining spectra were retrieved from the ESO Science 
Archive\footnote{http://archive.eso.org/eso/eso\_archive\_main.html}.

The data reduction was performed using a modified version of the FEROS-DRS pipeline (running under the ESO-MIDAS context FEROS) 
through the following reduction steps: bias subtraction and bad-column masking; definition of the \'echelle orders on flat-field 
frames; subtraction of the background diffuse light; order extraction; order-by-order flat-fielding; determination of wavelength-dispersion 
solution by means of ThAr calibration lamp exposures; order-by-order normalisation; re-binning to a linear wavelength-scale with barycentric 
correction; merging of the \'echelle orders. In the end, the final signal-to-noise ($S/N$) ratio of the wavelength-calibrated, merged, 
normalised spectra is in the range 80--250.

The FEROS spectra cover the range 3600--9200~\AA\ at the resolution $R=48\,000$. This wide spectral range allowed us to select 125+11 
Fe~{\sc i}+Fe~{\sc ii} lines, as well as spectral features of $\alpha$, iron-peak, and other elements (see Sect. \ref{sec:abundance}).

\section{Abundance measurements}
\label{sec:abundance}
\subsection{Lithium}
Lithium equivalent widths (EWs) were measured by direct integration or by deblending the observed line profiles using the 
IRAF task {\sc splot}. Then, lithium abundances, $\log n{\rm (Li)}$, were derived by interpolating the curves-of-growth of 
\cite{soderblometal1993} at the stellar $T_{\rm eff}$ and $\log g$ determined spectroscopically as described in Section~\ref{sec:parameters}.

\subsection{Iron-peak, $\alpha$, and other elements}
\label{sec:other_elements}
Elemental abundances of Na, Mg, Al, Si, Ca, Ti, Cr, Fe, Ni, and Zn were derived from the measurements of line EWs (see 
Section~\ref{sec:lin_sol_ew}) using the 2010 version of MOOG (\citealt{sneden1973}) and assuming local thermodynamic equilibrium (LTE). 
Radiative and Stark broadenings are treated in a standard way in MOOG (\citealt{barklemomara1997}), while for collisional broadening we used 
the \cite{unsold1955} approximation. \cite{Kuru93} grids of plane-parallel model atmospheres were used. 

\subsection{Line list, solar analysis, and equivalent widths}
\label{sec:lin_sol_ew}
We adopted the line list of Biazzo et al. (2011a, and references therein) integrated with lines from other lists 
(\citealt{clementinietal2000,bensbyetal2003,dorazirandich2009,dorazietal2009}). We refer to those papers for details on atomic parameters 
and their sources. The complete line list is given in Table~A1.

Our analysis was performed differentially with respect to the Sun. We analysed a solar (sky) spectrum acquired with FEROS, 
using our line list and the solar parameters ($T_{\rm eff}=5770$ K, $\log g=4.44$, $\xi=1.1$ km s$^{-1}$; 
see \citealt{randichetal2006, biazzoetal2011a}), 
and obtained $\log n$(Fe {\sc i})$_{\odot}$=7.50$\pm$0.05 and $\log n$(Fe {\sc ii})$_{\odot}$=7.50$\pm$0.06. 
With the aforementioned solar parameters, $\log n$(Fe {\sc i})$_{\odot}$ vs. EW and $\chi$ did not show any significant 
trends, implying that the assumed effective temperature and microturbulence represent quite well the corresponding real solar values.
The results for all the elements are reported in Table~\ref{tab:solar_abun} together with those given by \cite{grevesseetal1996} and 
\cite{asplundetal2009}. The latter values were obtained using 3D models. Our determinations are in good agreement with those from the 
literature (Table~\ref{tab:solar_abun}).

\begin{table}
 \centering
  \caption{Comparison between solar abundances we derived using Kurucz (1993) model atmospheres and the standard values from 
  Grevesse et al. (1996) and Asplund et al. (2009).\label{tab:solar_abun}}
  \begin{tabular}{@{}lccc@{}}
  \hline
Element     & $\log n_{\rm ATLAS}$ & $\log n_{\rm G96}$& $\log n_{\rm AS09}$\\
  \hline
 Na~{\sc i} & 6.36$\pm$0.07 & 6.33$\pm$0.03 & 6.24$\pm$0.04\\
 Mg~{\sc i} & 7.53$\pm$0.09 & 7.58$\pm$0.05 & 7.60$\pm$0.04\\
 Al~{\sc i} & 6.48$\pm$0.06 & 6.47$\pm$0.07 & 6.45$\pm$0.03\\
 Si~{\sc i} & 7.59$\pm$0.04 & 7.55$\pm$0.05 & 7.51$\pm$0.03\\
 Ca~{\sc i} & 6.35$\pm$0.08 & 6.36$\pm$0.02 & 6.34$\pm$0.04\\
 Ti~{\sc i} & 4.97$\pm$0.06 & 5.02$\pm$0.06 & 4.95$\pm$0.05\\
 Ti~{\sc ii}& 4.97$\pm$0.10 &             &  \\
 Cr~{\sc i} & 5.63$\pm$0.04 & 5.67$\pm$0.03 & 5.64$\pm$0.04\\
 Fe~{\sc i} & 7.50$\pm$0.05 & 7.50$\pm$0.04 & 7.50$\pm$0.04\\
 Fe~{\sc ii}& 7.50$\pm$0.06 &             &        \\
 Ni~{\sc i} & 6.26$\pm$0.07 & 6.25$\pm$0.01 & 6.22$\pm$0.04\\
 Zn~{\sc i} & 4.52$\pm$0.01 & 4.60$\pm$0.08 & 4.56$\pm$0.05\\
\hline
\end{tabular}
\end{table}

The line EWs of the target stars were measured using the automatic code ARES\footnote{http://www.astro.up.pt/$\sim$sousasag/ares/} 
(\citealt{sousaetal2007}). Very strong lines ($EW\gtsim150$ m\AA), which are heavily affected by the treatment of damping, 
were excluded; furthermore, a 2-$\sigma$ clipping criterion was applied to the initial Fe~{\sc i} line list before 
determining stellar parameters (Section~\ref{sec:parameters}) and iron abundance. Thus, for a given star, lines from the initial 
line list having a dispersion larger than a factor of two the {\it rms} were excluded.
Abundances of other elements were derived using the same criteria.

\subsection{Stellar parameters: effective temperature, micro-turbulence velocity, and surface gravity}
\label{sec:parameters}

Initial effective temperatures $T_{\rm eff}$ were set to the values obtained from the spectral types listed in Table~\ref{tab:literature} 
by applying the \cite{kenyonhartmann1995} calibrations. Then, final effective temperatures were determined by imposing the condition 
that the abundance from Fe~{\sc i} lines does not depend on the excitation potential of the lines. These temperatures are reported 
in Table~\ref{tab:param_chem} and represent the values adopted for the abundance analysis.

To infer the micro-turbulence velocity $\xi$, we first assumed 1.5 km\,s$^{-1}$ as initial value, and then imposed that the abundance from 
Fe~{\sc i} lines was independent on line EWs. Final values of $\xi$ are listed in Table~\ref{tab:param_chem}. As also found 
by \cite{padgett1996} and \cite{jamesetal2006}, the derived microturbulence is higher than the values for main sequence dwarfs 
of similar temperature. As stressed by \cite{santosetal2008}, the cause for this behavior is still unclear, but it may be related to 
chromospheric activity (\citealt{steenbockholweger1981}).

We estimated the surface gravity $\log g$ by imposing the Fe~{\sc i}/Fe~{\sc ii} ionisation equilibrium. The initial value was 
set to $\log g=4.4$ (e.g., almost the solar value). Final values of $\log g$ are listed in Table~\ref{tab:param_chem}. For comparison, 
we have computed the stellar surface gravity using the following relationship: 
$\log g=\log g_\odot + \log (M/M_\odot)+ 4 \log (T_{\rm eff}/T_{\rm eff}^\odot) - \log(L/L_\odot)$, where a solar gravity of 
4.44 dex, a solar effective temperature of 5770 K, a solar bolometric magnitude of 4.64 mag (\citealt{cox2000}), and a relation 
$L/L_\odot=(M/M_\odot)^{3.5}$ between stellar luminosity and mass were adopted. We have verified that our final values (listed in 
Table~\ref{tab:param_chem}, third column) are in good agreement (within 0.1 dex) with those derived through parallax measurements.

\subsection{Error budget}
\label{sec:abun_errors}

The quality of the measured line EWs depends on the spectral resolution, the $S/N$ ratio of the spectrum, the definition of the photospheric 
continuum adjacent to the line, and the projected rotational velocity of the star. The high-resolution, and high $S/N$ of the spectra used 
in the present analysis allowed us to measure lithium EWs in a very accurate way. Uncertainties in the lithium abundance derived through 
curves-of-growth are assessed by varying the input parameters, i.e. the EWs, 
effective temperatures, and surface gravities within their error bars. Considering the typical uncertainties of $\pm 5$ m\AA\,in 
lithium EWs, of $\pm 60$ K in $T_{\rm eff}$, and of $0.1$~dex in $\log g$ (see below), the resulting total error in $\log n{\rm (Li)}$ 
amounts typically to less than $\sim 0.10$~dex. Morever, in three stars (namely, HIP~82688, TYC~5155-1500-1, and HIP~27072), 
the Li~{\sc i} $\lambda$6707.8 m\AA~line is blended with the Fe~{\sc i} $\lambda$6707.4 m\AA~line, leading to an overestimation 
of the lithium EWs. Taking advantage of the empirical correction to the lithium line reported by \cite{soderblometal1993}, we 
estimated $0.02-0.06$~dex (i.e. $1-2$\%) as the uncertainty in the lithium abundance due to the iron contribution; this contribution 
is negligible when compared to the other error sources.

Elemental abundances of all other elements are affected by random (internal; $i.$) and systematic (external; $ii.$) errors. $i.$ Sources of 
internal errors include uncertainties in atomic and stellar parameters, and measured EWs. Uncertainties in atomic 
parameters, such as the transition probability ($\log gf$), should cancel out, since our analysis is carried out differentially with respect 
to the Sun. Errors due to uncertainties in stellar parameters ($T_{\rm eff}$, $\xi$, $\log g$) were estimated first by assessing errors 
in stellar parameters themselves and then by varying each parameter separately, while keeping the other two unchanged. We found that variations 
in $T_{\rm eff}$ larger than 60~K would introduce spurious trends in $\log n{\rm (Fe)}$ versus the excitation potential ($\chi$), while 
variations in $\xi$ larger than 0.1 km s$^{-1}$ would result in significant trends of $\log n{\rm (Fe)}$ versus EW, and variations in $\log g$ 
larger than 0.1 dex would lead to differences between $\log n$(Fe~{\sc i}) and $\log n$(Fe~{\sc ii}) larger than 0.05 dex. The above values were 
thus assumed as uncertainties in stellar parameters. Errors in abundances (both [Fe/H] and [X/Fe]) due to uncertainties in stellar parameters 
are summarised in Table~\ref{tab:errors} for the coolest (HD\,38392) and one of the warmest (HIP\,27072) stars in our sample. As for the 
errors due to uncertainties in EWs, our spectra are characterised by different $S/N$ ratios. As a consequence, random errors in EW are well 
represented by the standard deviation around the mean abundance determined from all the lines. These errors are listed in 
Table~\ref{tab:param_chem}, where uncertainties in [X/Fe] were obtained by quadratically adding the [Fe/H] error and the [X/H] error. 
When only one line was measured, the error in [X/H] is the standard deviation of three independent EW measurements. The number of lines 
employed for the abundance analysis is listed in Table~\ref{tab:param_chem} in parentheses. $ii.$ External or systematic errors, originated 
for instance by the code and/or the model atmospheres, should not influence largely our final abundance measurements (see \citealt{biazzoetal2011a}, 
and references therein). 

\begin{table*}  
\caption{Internal errors in abundance determination due to uncertainties in stellar parameters for the coolest 
star (namely, HD\,38392) and for one of the warmest (namely, HIP\,27072) in our sample. Numbers refer to the differences 
between the abundances obtained with and without the uncertainties in stellar parameters.}
\label{tab:errors}
\begin{center}
\begin{tabular}{lccc}
\hline
\hline
HD\,38392 & $T_{\rm eff}=5100$ K & $\log g=4.6$ & $\xi=1.5$ km/s\\
\hline
$\Delta$   & $\Delta T_{\rm eff}=-/+60$ K & $\Delta \log g=-/+0.1$ & $\Delta \xi=-/+0.1$ km/s\\
\hline
$[$Fe~{\sc i}/H$]$  & $-$0.02/0.01 & $-$0.01/0.00 & 0.01/$-$0.02  \\
$[$Fe~{\sc ii}/H$]$ & 0.06/$-$0.05 & $-$0.08/0.07 & 0.01/$-$0.01     \\
$[$Na/Fe$]$         & $-$0.02/0.03 & 0.02/$-$0.01 & 0.00/0.01	  \\
$[$Mg/Fe$]$         & 0.01/0.00 & 0.02/$-$0.02 & 0.00/0.00     \\
$[$Al/Fe$]$         & $-$0.01/0.02 & 0.02/$-$0.01 & 0.00/0.01 \\
$[$Si/Fe$]$         & 0.05/$-$0.03 & $-$0.01/0.03 & 0.00/0.02  \\
$[$Ca/Fe$]$         & $-$0.04/0.04 & 0.04/$-$0.03 & 0.01/0.00	  \\
$[$Ti~{\sc i}/Fe$]$ & $-$0.06/0.06 & 0.02/$-$0.01 & 0.02/$-$0.02  \\
$[$Ti~{\sc ii}/Fe$]$& 0.04/$-$0.03 & $-$0.05/0.05 & 0.00/0.01  \\
$[$Cr/Fe$]$         & $-$0.03/0.04 & 0.03/$-$0.01 & 0.02/0.00  \\			       
$[$Ni/Fe$]$         & 0.02/$-$0.01 & $-$0.01/0.02 & 0.01/0.00  \\				 
$[$Zn/Fe$]$         & 0.04/$-$0.03 & $-$0.03/0.02 & 0.01/$-$0.01  \\				       
\hline	\\	
HIP\,27072 & $T_{\rm eff}=6350$ K & $\log g=4.3$ & $\xi=1.6$ km/s\\
\hline
$\Delta$   & $\Delta T_{\rm eff}=-/+60$ K & $\Delta \log g=-/+0.1$ & $\Delta \xi=-/+0.1$ km/s\\
\hline
$[$Fe~{\sc i}/H$]$  & $-$0.04/0.04 & 0.00/0.00 & 0.01/$-$0.02  \\
$[$Fe~{\sc ii}/H$]$ & 0.01/$-$0.01 & $-$0.04/0.04 & 0.02/$-$0.02    \\
$[$Na/Fe$]$         & 0.01/$-$0.01 & 0.01/0.00 & 0.00/0.02	  \\
$[$Mg/Fe$]$         & 0.01/$-$0.01 & 0.01/$-$0.01 & 0.01/0.01     \\
$[$Al/Fe$]$         & 0.01/$-$0.02 & 0.00/0.00 & $-$0.01/0.01  \\
$[$Si/Fe$]$         & 0.02/$-$0.02 & 0.00/0.00 & 0.00/0.02  \\
$[$Ca/Fe$]$         & 0.00/0.00 & 0.02/$-$0.01 & 0.02/0.00     \\
$[$Ti~{\sc i}/Fe$]$ & $-$0.01/0.01 & 0.00/0.00 & $-$0.01/0.01  \\
$[$Ti~{\sc ii}/Fe$]$& 0.04/$-$0.04 & $-$0.04/0.04 & 0.00/0.01  \\
$[$Cr/Fe$]$         & $-$0.01/0.00 & 0.00/0.00 & 0.00/0.00  \\  			   
$[$Ni/Fe$]$         & 0.00/0.00 & 0.00/0.00 & 0.01/0.01  \\				   
$[$Zn/Fe$]$         & 0.01/$-$0.01 & 0.00/0.01 & 0.03/$-$0.01  \\  			      
\hline	\\	
\end{tabular}
\end{center}
\end{table*}

\section{Results}
\label{sec:results}

\subsection{Lithium abundance}
\begin{figure}	
\begin{center}
 \begin{tabular}{c}
 \hspace{-.6cm}
\includegraphics[width=9cm]{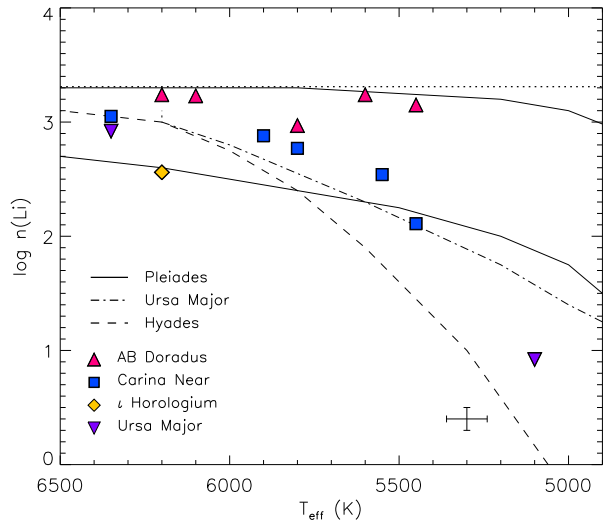}
 \end{tabular}
\caption{Lithium abundance versus effective temperature for all our sample. Typical errors in $T_{\rm eff}$ and $\log n{\rm (Li)}$ are 
over-plotted on the bottom-right corner of the panel. The dotted line at $\log n{\rm (Li)}=3.31$ marks the meteoritic abundance of lithium. 
The two solid lines represent the lower and upper limits for the Pleiades stars, while dash-dotted and dashed lines indicate the upper limits 
for the UMa and Hyades groups, respectively, as adopted by Zickgraf et al. (2005; see also references therein).
}
\label{fig:lithium_abund_associations}
 \end{center}
\end{figure}

In Fig.~\ref{fig:lithium_abund_associations}, we show the lithium abundance versus the spectroscopic effective temperature 
listed in Table~\ref{tab:param_chem}.

The members of the Ursa Major and Carina Near groups lie between the lower and upper envelopes of the Pleiades stars, as also found by 
\cite{zuckermanetal2006}, and close to the UMa upper envelope, with the exception of HD\,38392 which is slightly below the UMa envelope at 
cooler temperatures. The similarity between our UMa and Carina Near sample in the $\log n{\rm (Li)}-T_{\rm eff}$ diagram could be an indication 
of similar ages among the clusters. On the other hand, the AB~Dor members show mean lithium abundance of $\sim 3.2$ dex, without evidence of 
decreasing trend with temperature. Their position is close to the Pleiades upper envelope, confirming their younger age when compared to the 
other associations. Finally, $\iota$\,Hor shows $\log n{\rm (Li)}=2.56$ dex, confirming the older age of the star as compared 
with the other clusters (see Section~\ref{sec:iota_Hor}). Its $\log n{\rm (Li)}$ value is slightly below the Hyades upper envelope and close 
to the Pleiades lower envelope at its effective temperature. 

\subsection{Abundances of iron-peak, $\alpha$-, and other elements}
\label{sec:iron}
\subsubsection{Fe}
\label{sec:fe}

In Fig.~\ref{fig:all_abund_associations} we show the iron abundance ([Fe/H]) as a function of $T_{\rm eff}$ for the three associations 
and for $\iota$\,Hor. Since we obtain similar Fe~{\sc i} and Fe~{\sc ii} abundances for the whole sample (see Table~\ref{tab:param_chem} 
and top-left panel in Fig.~\ref{fig:all_abund_associations}), henceforth we will consider [Fe~{\sc i}/H] as iron abundance.

For the AB~Dor group we derive an average iron abundance of [Fe/H]=$0.10\pm0.03$ dex, which is well in agreement with the value of 
[Fe/H]=$0.04\pm0.05$ dex reported by \cite{vianaetal2009}. In particular, for the two stars in common with us, the differences 
between our values and theirs are: $\Delta T_{\rm eff}=$69 K, $\Delta \log g=$0.07 dex, $\Delta \xi=$0.27 km~s$^{-1}$, 
$\Delta$[Fe/H]=0.06 dex (TYC\,9493-838-1) and $\Delta T_{\rm eff}=$50 K, $\Delta \log g=$0.10 dex, $\Delta \xi=$0.18 km~s$^{-1}$, 
$\Delta$[Fe/H]=$-0.02$ dex (HIP\,114530). The (small) differences are within the uncertainties and they can be attributed to 
the different line lists and $\sigma$-clipping criteria used. In addition, the mean iron abundance we find for AB~Dor (e.g., 
[Fe/H]=$0.10\pm0.03$) is slightly larger than that of the Pleiades ([Fe/H]=$0.04\pm0.03$; \citealt{anetal2007}, and references therein). 
However, considering possible systematic differences between abundance analysis performed in different way, this does not role out 
the direct link between AB~Dor and Pleiades discussed by \cite{ortegaetal2007}. 

The Carina Near group shows a mean iron abundance of [Fe/H]=$0.08\pm0.06$ dex. For the two stars in common with \cite{desideraetal2006b}, 
we find similar values both in stellar parameters and in [Fe/H]. 

For the UMa members, we obtain a mean iron abundance of [Fe/H]=$0.03\pm0.02$ dex, which is close to recent results obtained through 
similar spectroscopic methods (see, e.g., the recent findings by \citealt{paulsonyelda2006}). Despite the low statistics, we can 
cautiously highlight the small abundance scatter of the UMa group, as also claimed in a recent work (\citealt{ammler-guenther2009}). 
We find that the (solar) iron abundance of the UMa group is very close to that of the Pleiades ([Fe/H]=$0.04\pm0.03$; \citealt{anetal2007}, 
and references therein). In particular, the UMa stars in our sample were recently analysed by \cite{paulsonyelda2006} and 
\cite{ramirezetal2007} through spectroscopic methods similar to ours, and all results agree within the errors.

The planet-host star $\iota$\,Hor shows [Fe/H]=$0.16\pm0.09$ dex. The case of this star will be discussed in Section \ref{sec:iota-hor}.

\subsubsection{Mg, Si, Ca, and Ti}
\label{sec:mg_si_ca_ti}
The $\alpha$-elements, such as magnesium, silicon, calcium, and titanium, are primarily produced in the aftermath of explosions of type 
II supernovae, with a small contribution from type Ia SNe (\citealt{woosleyweaver1995}).

\begin{figure*}	
\begin{center}
 \begin{tabular}{c}
\includegraphics[width=18cm]{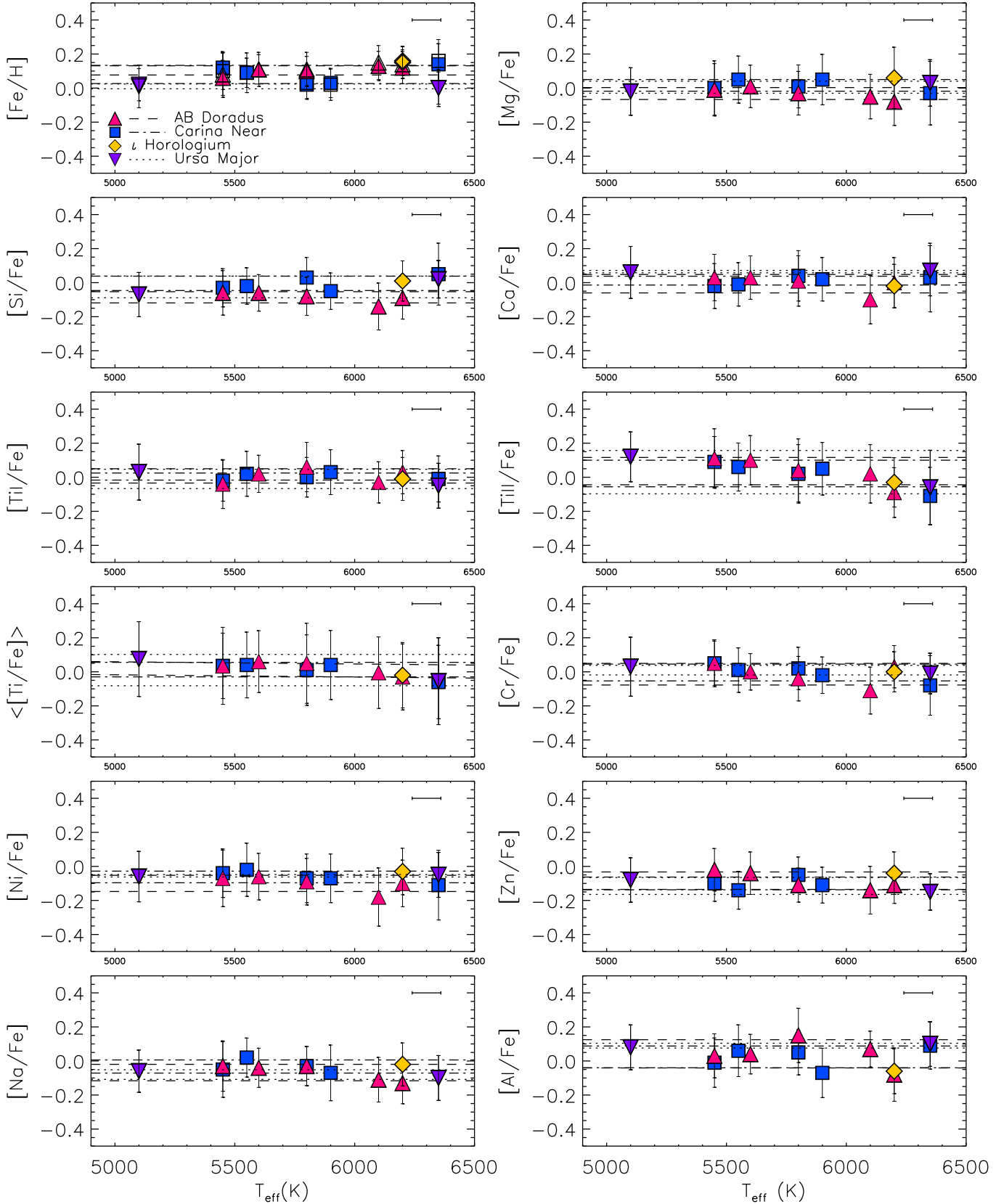}
 \end{tabular}
\caption{[Fe/H] and [X/Fe] versus $T_{\rm eff}$. In the [Fe/H] plot, filled and empty symbols refer to Fe~{\sc i} and Fe~{\sc ii} 
abundances, respectively. The meaning of the symbols is explained in the top-left panel. Different lines refer to regions of 
$\pm 1\sigma$ around the average, as indicated in the top-left panel. The horizontal error bar drawn in all plots 
represents the typical uncertainty in $T_{\rm eff}$.}
\label{fig:all_abund_associations}
 \end{center}
\end{figure*}

The abundances of $\alpha$-elements are listed in Table~\ref{tab:param_chem} and plotted in Fig.~\ref{fig:all_abund_associations}. 
The figure shows that there is no star-to-star variation for the different elements, which show solar [X/Fe] values, with the only possible 
exception of Ti~{\sc ii}, for which slight NLTE effects may be present (see \citealt{dorazirandich2009}, \citealt{biazzoetal2011a,biazzoetal2011b}, 
for thorough discussions on this issue). Therefore, we consider as titanium abundance the one obtained from Ti~{\sc i}.

The average silicon abundance we find for AB~Dor is in good agreement with the results of \cite{vianaetal2009}, who derived mean 
[Si/Fe]$=-0.07\pm0.05$ dex. In particular, for the two stars in common, the mean difference is only $0.02\pm0.04$ dex.

In $\iota$\,Hor, the abundance ratios of $\alpha$-elements with respect to iron are in their solar proportions, as also found by 
\cite{paulsonetal2003} for Hyades F--K dwarfs.

\subsubsection{Cr and Ni}
\label{sec:cr_ni}
Iron-peak elements are synthesised by SNIa explosions. In particular, Cr varies tighly in lockstep with iron at all [Fe/H], while 
Ni seems to show an upturn at [Fe/H]$>0$ (\citealt{bensbyetal2003}).

We measured the abundances of Cr and Ni as iron-peak elements; their values are plotted in Fig.~\ref{fig:all_abund_associations} 
as a function of $T_{\rm eff}$. Also in this case, all [X/Fe] values are consistent with the solar abundances, with very small scatter 
(Table~\ref{tab:param_chem}).

The average nickel abundance we derive for the AB~Dor group is in good agreement with the results by \cite{vianaetal2009}, i.e. 
$<$[Ni/Fe]$>=-0.06\pm0.10$ dex. In particular, for the two stars in common with us, the mean difference is only $-0.01\pm0.06$ dex.

\subsubsection{Zn}
\label{sec:zn}
Zinc is a volatile element which appears to behave similarly to the $\alpha$-elements. 

The [Zn/Fe] ratio is slightly lower than the solar value for all associations, while for $\iota$\,Hor we find solar abundance, 
in agreement with \cite{paulsonetal2003} for Hyades F--K dwarfs. 

\subsubsection{Na and Al}
\label{sec:na_al}
Sodium and aluminium are thought to be produced in SNe II and SNe Ib/c (\citealt{nomotoetal1984}) as a consequence of Ne and Mg burnings 
in massive stars through NeNa and MgAl chains. 

We plot the abundances of Na and Al relative to Fe versus $T_{\rm eff}$ in Fig.~\ref{fig:all_abund_associations}. The abundance 
ratios of Na and Al of the studied sample with respect to Fe are in their solar proportions (similar result was found by 
\citealt{paulsonetal2003} for the sodium abundance of Hyades dwarfs). 

\section{Discussion}
\label{sec:discussion}

\subsection{Elemental abundances of young associations in the Galactic disc}
\label{sec:galactic_disc}
Each component of the Milky Way (bulge, halo, thin/thick disc) presents a characteristic elemental abundance pattern, whose differences 
reflect a variety of star formation histories. In our case, the small elemental abundance dispersion of the three associations studied 
here agrees with other recent investigations conducted both in star-forming regions (e.g., \citealt{santosetal2008,biazzoetal2011a}) 
and in nearby young associations (e.g., \citealt{vianaetal2009,ammler-guenther2009}). When compared with local field stars of the 
thin Galactic disc (\citealt{soubirangirard2005}), AB\,Doradus, Carina Near, and Ursa Major show a similar abundance pattern. 
This suggests that the gas from which they formed did not undergo a peculiar enrichment, confirming the findings by Biazzo et al. 
(2011b; see their Fig.~10). This suggests that these nearby associations are representative of the current abundance 
(in all elements analysed here) of the Galactic thin disc in the solar neighbourhood.

\subsection{Are nearby young associations good candidates to search for exo-planets?}
\label{sec:exoplanets}
The results of abundance analysis of young stars show that none of the young associations or moving groups with available metallicity 
determination is extremely metal-rich (see, e.g., \citealt{santosetal2008, vianaetal2009, biazzoetal2011a, biazzoetal2011b, dorazietal2011}; 
and this work). Members of nearby young moving groups are the best targets for planet searches using direct imaging techniques. 
Surveys focusing on these targets were performed in the past years (e.g., \citealt{kasperetal2007, chauvinetal2010}) and next generation 
direct imaging instruments, like SPHERE and GPI\footnote{Gemini Planet Imager}, will also intensively observe these stars. 
However, it is emerging that the metallicity distribution of nearby young stars studied in direct imaging surveys is different from that 
of samples of radial velocity survey (e.g., \citealt{fischervalenti2005}). Such metallicity distribution is characterised by a significantly 
lower dispersion and a slightly lower mean metallicity. As suggested by \cite{liviopringler2003}, metallicity may play an impotant role 
in the migration history of planets. Thus, the lack of nearby, young super-metal-rich stars complicates the comparison of the results 
of planet searches around these stars with those from radial velocity surveys. First of all, because the low metallicity dispersion 
of young stars\footnote{The $\sim$30 Myr old planet host HR~8799 was found to have sub-solar metallicity ([Fe/H]=$-$0.5), but 
with the abundance pattern typical of $\lambda$~Boo stars (\citealt{sadakane2006}), in which the low metallicity is usually ascribed 
to the details of the star's accretion and atmospheric physics rather than an initial low metallicity of the system 
(\citealt{graycorbally2002}).} makes it challenging to investigate any trend in the frequency of giant planets at wide separations with 
metallicity. Second, because the statistical interpretation of the results of direct imaging surveys is often done by comparing the observed 
frequencies or upper limits with extrapolations of the results of radial velocity surveys.

The slightly sub-solar metallicity of nearby young associations cannot be explained in terms of radial Galactocentric migration, as they are 
younger than $\sim 600$ Myr. In the scenario devised by \cite{haywood2009}, giant planet formation could be favored 
at Galactocentric radii where the density of the molecular hydrogen, the primary constituent of planets, is higher (in particular, at 
the position of the molecular ring). In this case, one might expect a paucity of giant planets around young stars as compared to older 
stars, originating closer to the Galactic centre.

\subsection{The case of $\iota$\,Hor: a metal-rich planet-host star in the Hyades stream}
\label{sec:iota-hor}
For this target, we derive $T_{\rm eff}$=6200 K, $\log g=$4.5, [Fe~{\sc i}/H]=0.16$\pm$0.09 dex, [Fe~{\sc ii}/H]=0.15$\pm$0.09 dex, 
and [X/H](=[X/Fe]+[Fe/H]) higher than the Sun (see Table~\ref{tab:param_chem}). Our values of astrophysical parameters 
and elemental abundances are in very good agreement with the recent literature values listed in Table~\ref{tab:literature_iota_hor}, 
with the only exception of \cite{bondetal2006}'s results, who found lower [Fe/H]. This agreement confirms that this star is more metal-rich 
than the Sun at the $\sim 2 \sigma$ level. Although this is still marginally consistent with what is expected from statistical fluctuations, 
some discussion of this object would be merited. 

As we mentioned in the Introduction, $\iota$\,Hor is a young planet-host star most probably formed within the primordial Hyades star cluster 
and then evaporated towards the present position. Its iron abundance is very close to the value of the Hyades cluster (i.e. 0.13$\pm$0.05; 
\citealt{paulsonetal2003}), but also its [X/Fe] are in agreement with the mean cluster values (i.e. $<$[Na/Fe]$>$=0.01$\pm$0.09, 
$<$[Mg/Fe]$>$=$-$0.06$\pm$0.04, $<$[Si/Fe]$>$=0.05$\pm$0.05, $<$[Ca/Fe]$>$=0.07$\pm$0.07, $<$[Ti/Fe]$>$=0.03$\pm$0.05, 
$<$[Zn/Fe]$>$=$-$0.06$\pm$0.06; \citealt{paulsonetal2003}). This supports the idea that the origin of the over-metallicity and over-abundance 
in all other elements (i.e. [X/H]) is primordial, and not due to planet accretion, i.e. $\iota$\,Hor seems to be formed together with the 
other Hyades stars, at the same time and in the same primordial cloud. This will have important implications both for theories of 
star/exoplanet formation and for cluster formation and evolution (\citealt{vauclairetal2008}).

\section{Conclusions}
\label{sec:conclusions}
In this paper, we presented abundance measurements of iron-peak elements, $\alpha$-elements, and other odd-Z and even-Z elements 
in three young nearby associations (AB\,Doradus, Carina Near, and Ursa Major) and in the giant-planet host $\iota$\,Horologii. 
To this aim, we used FEROS high-resolution spectra. Our main results can be summarised as follows: 
\begin{enumerate}
\item Lithium abundance of all stars is consistent with their age. 
\item The three associations AB~Doradus, Carina Near, and Ursa Major have mean iron abundances of [Fe/H]=$0.10\pm0.03$, $0.08\pm0.05$, 
and $0.01\pm0.01$, respectively (where the error is the standard deviation on the average). These associations are characterised by small 
scatter in all elemental abundances.
\item The distribution of elemental abundances of the three associations is consistent with the thin disc population of the Galaxy.
\item For $\iota$\,Horologii, we find [Fe/H]=$0.16\pm0.09$ confirming its metal-richness. 
\item None of the members of the three nearby associations considered here is found to be super metal-rich (i.e. with highly super-solar 
metallicity). This confirms the general property of young nearby stars, i.e. that their metallicity differs from that of old, field, 
solar-type stars, which represent so far the most extensively planet-surveyed sample by radial velocity studies.
This fact will have necessarily to be taken into account for a proper interpretation of the results of direct-imaging planet searches, 
whose primary targets will be young nearby stars.
\end{enumerate}

\section*{Acknowledgements}
The authors are very grateful to the referee for a careful reading of the paper and for his/her comments that helped improving the paper. 
This paper makes use of data collected for the preparation of the SPHERE@ESO GTO survey. We warmly thank the SPHERE Consortium 
for making them available for the present work. This paper is also based on observations made with European Southern Observatory telescopes 
(program IDs: 70.D-0081(A), 082.A-9007(A), 083.A-9011(B), 084.A-9011(B)) and data obtained from the ESO Science Archive 
Facility under request numbers: 143106, 143382, 147882, 152598, 153529, 162614, 6572, 10960. This research made use of the 
SIMBAD database, operated at the CDS (Strasbourg, France). KB acknowledges the financial support from the INAF Post-doctoral fellowship. 
SD and EC acknowledge the PRIN-INAF 2010 ``Planetary systems at young ages and interactions with their active host stars''.

\newpage
\topmargin 6.5 cm
\pagestyle{empty}

\begin{landscape}
\begin{table*}
\tiny
\centering
\setlength{\tabcolsep}{1.4pt}
  \caption{Astrophysical parameters and elemental abundances derived from our analysis.\label{tab:param_chem}}
  \begin{tabular}{@{}lcccrrrrrrrrrrrrrcc@{}}
  \hline
\tiny{Star}&$\tiny{T_{\rm eff}}$&$\tiny{\log g}$&$\tiny{\xi}$&\tiny{[Fe~{\sc i}/H]}&\tiny{[Fe~{\sc ii}/H]}
&\tiny{[Na/Fe]}&\tiny{[Mg/Fe]}&\tiny{[Al/Fe]}&\tiny{[Si/Fe]}&\tiny{[Ca/Fe]}&\tiny{[Ti~{\sc i}/Fe]}
&\tiny{[Ti{\sc ii}/Fe]}&\tiny{$<$[Ti/Fe]$>$}$^{*}$&\tiny{[Cr/Fe]}&\tiny{[Ni/Fe]}&\tiny{[Zn/Fe]} &\tiny{$EW_{\rm Li}$}&\tiny{$\log n{\rm (Li)}$}\\ 
   &\tiny{(K)}  & & \tiny{(km/s)} & \tiny{(dex)} &\tiny{(dex)}& \tiny{(dex)}& \tiny{(dex)}& \tiny{(dex)}& \tiny{(dex)}& \tiny{(dex)}&\tiny{(dex)}& \tiny{(dex)}& \tiny{(dex)}& \tiny{(dex)}& \tiny{(dex)}& \tiny{(dex)}& (m\AA) &\tiny{(dex)}\\  
 \hline
\multicolumn{17}{c}{AB Doradus}\\
TYC~9493-838-1 &5450&4.6&1.8&0.06$\pm$0.10(117)&0.08$\pm$0.13(10)&$-$0.03$\pm$0.15(3)&$-$0.01$\pm$0.15(2)&   0.03$\pm$0.13(2)&$-$0.06$\pm$0.13(10)&   0.03$\pm$0.14(11)&$-$0.04$\pm$0.14(13)&	0.11$\pm$0.17(3)&   0.04$\pm$0.23&   0.05$\pm$0.13(11)&$-$0.07$\pm$0.17(35)&$-$0.02$\pm$0.12(1)&228&3.15\\
HIP~114530     &5600&4.6&1.6&0.11$\pm$0.09(118)&0.11$\pm$0.10(10)&$-$0.04$\pm$0.11(3)&   0.01$\pm$0.12(2)&   0.03$\pm$0.12(2)&$-$0.06$\pm$0.11(12)&   0.03$\pm$0.13(10)&   0.02$\pm$0.11(12)&   0.10$\pm$0.14(2)&   0.06$\pm$0.18&   0.00$\pm$0.11(10)&$-$0.06$\pm$0.14(38)&$-$0.04$\pm$0.12(1)&217&3.24\\
TYC~5155-1500-1&5800&4.6&1.9&0.10$\pm$0.09(108)&0.11$\pm$0.10(8) &$-$0.03$\pm$0.11(3)&$-$0.03$\pm$0.13(2)&   0.15$\pm$0.16(2)&$-$0.08$\pm$0.11(10)&   0.01$\pm$0.15(10)&   0.06$\pm$0.14(12)&	0.04$\pm$0.19(2)&   0.05$\pm$0.24&$-$0.04$\pm$0.13(3) &$-$0.09$\pm$0.14(34)&$-$0.11$\pm$0.10(1)&143&2.97\\
HIP~82688      &6100&4.6&1.9&0.13$\pm$0.09(54) &0.15$\pm$0.10(6) &$-$0.11$\pm$0.13(1)&$-$0.05$\pm$0.13(1)&   0.07$\pm$0.11(2)&$-$0.14$\pm$0.14(7) &$-$0.10$\pm$0.14(9) &$-$0.03$\pm$0.12(7) &	0.02$\pm$0.17(3)&   0.00$\pm$0.21&$-$0.11$\pm$0.14(10)&$-$0.18$\pm$0.17(23)&$-$0.14$\pm$0.14(1)&142&3.23\\
TYC~5901-1109-1&6200&4.6&1.5&0.12$\pm$0.09(117)&0.14$\pm$0.08(10)&$-$0.13$\pm$0.12(3)&$-$0.08$\pm$0.14(2)&$-$0.08$\pm$0.16(2)&$-$0.09$\pm$0.12(11)&   0.00$\pm$0.15(13)&   0.03$\pm$0.13(11)&$-$0.09$\pm$0.15(2)&$-$0.03$\pm$0.19&   0.03$\pm$0.12(11)&$-$0.10$\pm$0.14(34)&$-$0.11$\pm$0.11(1)&130&3.24\\
~\\
Average        &    &   &   &0.10$\pm$0.03     &0.12$\pm$0.03    &$-$0.07$\pm$0.05   &$-$0.03$\pm$0.03   &   0.04$\pm$0.08   &$-$0.09$\pm$0.03    &$-$0.01$\pm$0.05    &$-$0.03$\pm$0.04    &			&   0.02$\pm$0.04&$-$0.01$\pm$0.06    &$-$0.10$\pm$0.05    &$-$0.08$\pm$0.05&&\\
\hline
\multicolumn{17}{c}{Carina Near}\\
HIP~37923      &5450&4.6&1.5&0.12$\pm$0.09(113)&0.10$\pm$0.11(10)&$-$0.05$\pm$0.16(3)&   0.00$\pm$0.16(2)&$-$0.01$\pm$0.14(2)&$-$0.03$\pm$0.11(10)&$-$0.02$\pm$0.13(10)&$-$0.02$\pm$0.12(12)&	0.09$\pm$0.15(2)&   0.04$\pm$0.19&   0.05$\pm$0.14(10)&$-$0.04$\pm$0.14(37)&$-$0.10$\pm$0.11(1)&68 &2.11\\
HIP~37918      &5550&4.7&1.7&0.09$\pm$0.09(113)&0.09$\pm$0.12(10)&   0.02$\pm$0.11(3)&   0.05$\pm$0.14(2)&   0.06$\pm$0.15(2)&$-$0.02$\pm$0.11(10)&$-$0.01$\pm$0.13(10)&   0.02$\pm$0.13(12)&	0.06$\pm$0.14(2)&   0.04$\pm$0.19&   0.01$\pm$0.13(10)&$-$0.02$\pm$0.16(37)&$-$0.14$\pm$0.11(1)&114&2.54\\
HIP~58241$^{**}$&5800&4.6&1.7&0.02$\pm$0.07(99) &0.03$\pm$0.07(11)&$-$0.03$\pm$0.11(3)&   0.01$\pm$0.13(2)&   0.05$\pm$0.13(2)&   0.03$\pm$0.12(9) &   0.04$\pm$0.15(11)&   0.00$\pm$0.12(13)&	0.02$\pm$0.17(2)&   0.01$\pm$0.21&   0.02$\pm$0.12(11)&$-$0.07$\pm$0.14(37)&$-$0.05$\pm$0.11(1)&113&2.77\\
HIP~58240      &5900&4.6&1.6&0.03$\pm$0.09(114)&0.02$\pm$0.09(10)&$-$0.07$\pm$0.16(3)&   0.05$\pm$0.15(2)&$-$0.07$\pm$0.14(2)&$-$0.05$\pm$0.11(10)&   0.02$\pm$0.13(10)&   0.03$\pm$0.13(13)&	0.05$\pm$0.15(3)&   0.04$\pm$0.20&$-$0.02$\pm$0.11(11)&$-$0.07$\pm$0.14(36)&$-$0.11$\pm$0.11(1)&114&2.88\\
HIP~36414      &6350&4.5&2.4&0.14$\pm$0.12(55) &0.16$\pm$0.13(8) &      	     &$-$0.03$\pm$0.19(1)&   0.09$\pm$0.14(1)&   0.05$\pm$0.18(4) &   0.03$\pm$0.20(6) &$-$0.01$\pm$0.14(2) &$-$0.11$\pm$0.17(2)&$-$0.06$\pm$0.22&$-$0.08$\pm$0.17(3) &$-$0.11$\pm$0.20(9) & 	       	       &84 &3.05\\
~\\
Average        &    &   &   &0.08$\pm$0.05     &0.08$\pm$0.06	 &$-$0.03$\pm$0.04   &   0.02$\pm$0.03   &   0.02$\pm$0.06   &   0.00$\pm$0.04    &   0.01$\pm$0.03    &$-$0.02$\pm$0.02    &			&   0.01$\pm$0.04&   0.00$\pm$0.05    &$-$0.06$\pm$0.03    &$-$0.10$\pm$0.04&&\\
\hline
\multicolumn{17}{c}{Ursa Major}\\
HD~38392$^{**}$&5100&4.6&1.5&0.02$\pm$0.09(97) &0.01$\pm$0.13(8) &$-$0.06$\pm$0.12(1)&$-$0.02$\pm$0.14(1)&   0.08$\pm$0.13(2)&$-$0.07$\pm$0.13(9) &   0.06$\pm$0.15(7) &   0.03$\pm$0.16(12)&   0.12$\pm$0.15(1)&   0.08$\pm$0.22&   0.03$\pm$0.17(5) &$-$0.06$\pm$0.15(33)&$-$0.08$\pm$0.13(1) &10&0.92\\
HIP~27072$^{**}$&6350&4.3&1.6&0.00$\pm$0.09(75) &0.00$\pm$0.11(10)&$-$0.10$\pm$0.13(3)&   0.03$\pm$0.14(1)&   0.10$\pm$0.13(1)&   0.02$\pm$0.11(8) &   0.07$\pm$0.15(13)&$-$0.05$\pm$0.13(8) &$-$0.06$\pm$0.22(2)&$-$0.05$\pm$0.26&$-$0.01$\pm$0.12(10)&$-$0.05$\pm$0.13(30)&$-$0.15$\pm$0.11(1)&68&2.92\\
~\\
Average        &    &   &   &0.01$\pm$0.01     &0.01$\pm$0.01	 &$-$0.08$\pm$0.03   &   0.01$\pm$0.04   &   0.09$\pm$0.01   &$-$0.03$\pm$0.06    &   0.07$\pm$0.01    &$-$0.05$\pm$0.06    &			&   0.02$\pm$0.09&   0.01$\pm$0.03    &$-$0.05$\pm$0.01    &$-$0.12$\pm$0.05&&\\
\hline
\multicolumn{17}{c}{$\iota$\,Horologii (Hyades stream)}\\
HIP~12653      &6200&4.5&1.5&0.16$\pm$0.09(111)&0.15$\pm$0.09(10)&$-$0.02$\pm$0.13(3)&   0.06$\pm$0.18(1)&$-$0.06$\pm$0.13(2)&   0.01$\pm$0.12(10)&$-$0.02$\pm$0.13(11)&$-$0.01$\pm$0.13(12)&$-$0.03$\pm$0.14(3)&$-$0.02$\pm$0.19&   0.00$\pm$0.12(11)&$-$0.03$\pm$0.14(36)&$-$0.04$\pm$0.12(1)&38&2.56\\
\hline
\end{tabular}
$^{*}$ Average of [Ti{\sc i}/Fe] and [Ti{\sc ii}/Fe].\\
$^{**}$ The astrophysical parameters of these stars, also reported in \cite{dorazietal2012}, were revised. The results are slightly different but consistent within 
the errors.
\end{table*}
\end{landscape}
\normalsize

\topmargin 5 cm

\begin{landscape}
\begin{table*}
\tiny
\centering
\setlength{\tabcolsep}{1.8pt}
  \caption{Astrophysical parameters and elemental abundances of $\iota$\,Hor from the literature and from this work.\label{tab:literature_iota_hor}}
  \begin{tabular}{@{}lcccrrrrrrrrrrrrrcc@{}}
  \hline
$\tiny{T_{\rm eff}}$&$\tiny{\log g}$&$\tiny{\xi}$&\tiny{[Fe~{\sc i}/H]}&\tiny{[Fe~{\sc ii}/H]}
&\tiny{[Na/H]}&\tiny{[Mg/H]}&\tiny{[Al/H]}&\tiny{[Si/H]}&\tiny{[Ca/H]}&\tiny{[Ti~{\sc i}/H]}
&\tiny{[Ti{\sc ii}/H]}&\tiny{[Cr/H]}&\tiny{[Ni/H]}&\tiny{[Zn/H]} & Reference\\ 
   \tiny{(K)}  & & \tiny{(km/s)} & \tiny{(dex)} &\tiny{(dex)}& \tiny{(dex)}& \tiny{(dex)}& 
   \tiny{(dex)}& \tiny{(dex)}& \tiny{(dex)}&\tiny{(dex)}& \tiny{(dex)}& \tiny{(dex)}& \tiny{(dex)}&\tiny{(dex)}&\\  
 \hline
6150$\pm$70&&&0.14$\pm$0.10&0.12$\pm$0.08&0.13$\pm$0.04&0.15$\pm$0.06&0.13$\pm$0.09&0.17$\pm$0.09&0.23$\pm$0.14&0.14$\pm$0.15&0.14$\pm$0.18&0.16$\pm$0.13&0.12$\pm$0.11&0.05$\pm$0.18&\cite{bensbyetal2003}\\
6252$\pm$53&4.61$\pm$0.16&1.18$\pm$0.10&0.26$\pm$0.06&&&&&&&&&&&&\cite{santosetal2004}\\
&&&&&0.24$\pm$0.01&0.19$\pm$0.04&0.19$\pm$0.01&&&&&&&&\cite{beiraoetal2005}\\
6150&4.37&&0.14&&&0.148&&&&&&&&&\cite{borkovamarsakov2005}\\
6097 & 4.34 & & 0.11 & & 0.08 & &&0.10&&0.07&&&0.07&& \cite{fischervalenti2005}\\
6017$\pm$22&4.32$\pm$0.16&1.50$\pm$0.16&0.01$\pm$0.07&&0.06$\pm$0.05&&$-$0.17$\pm$0.05&0.09$\pm$0.07&0.07$\pm$0.13&$-$0.02$\pm$0.12&$-$0.03$\pm$0.06&0.03$\pm$0.04&&&\cite{bondetal2006}\\
&&&&&&&&0.19$\pm$0.05&0.17$\pm$0.11&0.26$\pm$0.07&&0.16$\pm$0.07&0.19$\pm$0.06&&\cite{gillietal2006}\\
&&&0.195$\pm$0.056&&0.208$\pm$0.038&0.175$\pm$0.055&0.167$\pm$0.036&0.158$\pm$0.051&0.103$\pm$0.100&0.160$\pm$0.070&&&0.142$\pm$0.054&&\cite{gonzalezlaws2007}\\
6227$\pm$26&4.53$\pm$0.06&1.29$\pm$0.03&0.19$\pm$0.02&&&&&&&&&&&&\cite{sousaetal2008}\\
&&&&&0.15$\pm$0.04&0.14$\pm$0.08&0.11$\pm$0.02&0.17$\pm$0.03&0.19$\pm$0.03&0.20$\pm$0.04&0.16$\pm$0.02&0.19$\pm$0.02&0.18$\pm$0.03&&\cite{nevesetal2009}\\
6080$\pm$80&4.40$\pm$0.06&0.15$\pm$0.07&&&&&&&&&&&&&\cite{brunttetal2010}\\
6200$\pm$60&4.5$\pm$0.1&1.5$\pm$0.1&0.16$\pm$0.09&0.15$\pm$0.09&0.14$\pm$0.09&0.21$\pm$0.16&0.10$\pm$0.10&0.17$\pm$0.08&0.14$\pm$0.09&0.15$\pm$0.09&0.13$\pm$0.12&0.16$\pm$0.08&0.13$\pm$0.11&0.12$\pm$0.09&This work\\
\hline
\end{tabular}
\end{table*}
\end{landscape}
\normalsize

\newpage
\topmargin 0 cm

\appendix

\section[]{Line list}

\begin{table}
\begin{center}
  \caption{Wavelength, elements, excitation potential, and oscillator strength. }
  \begin{tabular}{lllr}
  \hline
   $\lambda$ & Element & $\chi$ & $\log gf$ \\ 
   (\AA)     &         & (eV)   &           \\ 
 \hline
  5682.63    &  Na~{\sc i}  &  2.102 &     $-$0.700    \\
  6154.23    &  Na~{\sc i}  &  2.102 &     $-$1.610   \\
  6160.75    &  Na~{\sc i}  &  2.104 &     $-$1.310   \\
  7657.60    &  Mg~{\sc i}  &  5.110 &     $-$1.280   \\
  8310.30    &  Mg~{\sc i}  &  5.930 &     $-$1.090   \\ 
  6696.02    &  Al~{\sc i}  &  3.143 &     $-$1.499  \\
  6698.67    &  Al~{\sc i}  &  3.143 &     $-$1.950   \\
  5701.10    &  Si~{\sc i}  &  4.930 &     $-$2.050   \\
  5948.54    &  Si~{\sc i}  &  5.082 &     $-$1.230   \\
  6091.92    &  Si~{\sc i}  &  5.871 &     $-$1.400    \\
  6125.02    &  Si~{\sc i}  &  5.614 &     $-$1.570   \\
  6142.48    &  Si~{\sc i}  &  5.619 &     $-$1.480   \\ 
  6145.02    &  Si~{\sc i}  &  5.616 &     $-$1.440   \\ 
  6414.98    &  Si~{\sc i}  &  5.871 &     $-$1.100    \\  
  6518.73    &  Si~{\sc i}  &  5.954 &     $-$1.500    \\  
  6555.46    &  Si~{\sc i}  &  5.984 &     $-$1.000    \\  
  7003.58    &  Si~{\sc i}  &  5.960 &     $-$0.870   \\  
  7235.34    &  Si~{\sc i}  &  5.610 &     $-$1.510   \\  
  7918.40    &  Si~{\sc i}  &  5.950 &     $-$0.610   \\  
  7932.40    &  Si~{\sc i}  &  5.960 &     $-$0.470   \\  
  5512.98    &  Ca~{\sc i}  &  2.933 &     $-$0.480   \\  
  5581.97    &  Ca~{\sc i}  &  2.523 &     $-$0.671  \\  
  5601.28    &  Ca~{\sc i}  &  2.526 &     $-$0.523  \\  
  5867.56    &  Ca~{\sc i}  &  2.933 &     $-$1.610   \\  
  6102.72    &  Ca~{\sc i}  &  1.879 &     $-$0.862  \\  
  6122.22    &  Ca~{\sc i}  &  1.886 &     $-$0.386  \\  
  6161.30    &  Ca~{\sc i}  &  2.523 &     $-$1.293  \\  
  6166.44    &  Ca~{\sc i}  &  2.521 &     $-$1.156  \\  
  6169.04    &  Ca~{\sc i}  &  2.523 &     $-$0.804  \\  
  6169.56    &  Ca~{\sc i}  &  2.526 &     $-$0.527  \\  
  6455.60    &  Ca~{\sc i}  &  2.523 &     $-$1.424  \\  
  6499.65    &  Ca~{\sc i}  &  2.523 &     $-$0.818  \\  
  7148.15    &  Ca~{\sc i}  &  2.710 &     0.137   \\  
  7326.16    &  Ca~{\sc i}  &  2.930 &     $-$0.230   \\  
  4805.42    &  Ti~{\sc i}  &  2.345 &     0.150    \\  
  4820.41    &  Ti~{\sc i}  &  1.502 &     $-$0.441  \\  
  4885.08    &  Ti~{\sc i}  &  1.887 &     0.358   \\  
  4913.61    &  Ti~{\sc i}  &  1.873 &     0.160    \\  
  5016.16    &  Ti~{\sc i}  &  0.848 &     $-$0.574  \\  
  5219.70    &  Ti~{\sc i}  &  0.021 &     $-$2.292  \\  
  5866.45    &  Ti~{\sc i}  &  1.067 &     $-$0.840   \\  
  5953.16    &  Ti~{\sc i}  &  1.887 &     $-$0.329  \\  
  5965.83    &  Ti~{\sc i}  &  1.879 &     $-$0.409  \\  
  6258.10    &  Ti~{\sc i}  &  1.443 &     $-$0.431  \\  
  6261.10    &  Ti~{\sc i}  &  1.430 &     $-$0.479  \\  
  6743.13    &  Ti~{\sc i}  &  0.900 &     $-$1.630   \\  
  7357.74    &  Ti~{\sc i}  &  1.440 &     $-$1.120   \\  
  6491.56    &  Ti~{\sc ii} &  2.061 &     $-$1.793  \\  
  6606.95    &  Ti~{\sc ii} &  2.061 &     $-$2.790   \\  
  6680.13    &  Ti~{\sc ii} &  3.095 &     $-$1.855  \\  
  4936.34    &  Cr~{\sc i}  &  3.113 &     $-$0.340   \\  
  5247.57    &  Cr~{\sc i}  &  0.961 &     $-$1.640   \\  
  5296.69    &  Cr~{\sc i}  &  0.983 &     $-$1.400    \\  
\hline
\end{tabular}
\label{tab:line_list}
\end{center}
\end{table}

\begin{table}
\begin{center}
  \contcaption{}
  \begin{tabular}{lllr}
  \hline
   $\lambda$ & Element & $\chi$ & $\log gf$ \\ 
   (\AA)     &         & (eV)   &           \\ 
 \hline
  5300.74    &  Cr~{\sc i}  &  0.983 &     $-$2.120   \\  
  5329.14    &  Cr~{\sc i}  &  2.914 &     $-$0.064  \\  
  5348.31    &  Cr~{\sc i}  &  1.004 &     $-$1.290   \\  
  6883.07    &  Cr~{\sc i}  &  3.440 &     $-$0.420   \\  
  6925.28    &  Cr~{\sc i}  &  3.450 &     $-$0.200    \\  
  6926.10    &  Cr~{\sc i}  &  3.450 &     $-$0.590   \\  
  6979.80    &  Cr~{\sc i}  &  3.460 &     $-$0.220   \\  
  7355.90    &  Cr~{\sc i}  &  2.890 &     $-$0.290   \\  
  7400.19    &  Cr~{\sc i}  &  2.900 &     $-$0.110   \\  
  4835.87    &  Fe~{\sc i}  &  4.103 &     $-$1.500    \\  
  4875.88    &  Fe~{\sc i}  &  3.332 &     $-$2.020   \\  
  4907.73    &  Fe~{\sc i}  &  3.430 &     $-$1.840   \\  
  4999.11    &  Fe~{\sc i}  &  4.186 &     $-$1.740   \\  
  5036.92    &  Fe~{\sc i}  &  3.017 &     $-$3.068  \\  
  5044.21    &  Fe~{\sc i}  &  2.851 &     $-$2.059  \\  
  5067.15    &  Fe~{\sc i}  &  4.220 &     $-$0.930   \\  
  5141.74    &  Fe~{\sc i}  &  2.424 &     $-$2.190   \\  
  5162.27    &  Fe~{\sc i}  &  4.178 &     0.020    \\  
  5217.39    &  Fe~{\sc i}  &  3.211 &     $-$1.070   \\  
  5228.38    &  Fe~{\sc i}  &  4.220 &     $-$1.290   \\  
  5285.13    &  Fe~{\sc i}  &  4.434 &     $-$1.640   \\  
  5293.96    &  Fe~{\sc i}  &  4.143 &     $-$1.870   \\  
  5373.71    &  Fe~{\sc i}  &  4.473 &     $-$0.860   \\  
  5386.33    &  Fe~{\sc i}  &  4.154 &     $-$1.770   \\  
  5389.48    &  Fe~{\sc i}  &  4.415 &     $-$0.570   \\  
  5397.62    &  Fe~{\sc i}  &  3.634 &     $-$2.480   \\  
  5398.28    &  Fe~{\sc i}  &  4.445 &     $-$0.720   \\  
  5472.71    &  Fe~{\sc i}  &  4.209 &     $-$1.495  \\  
  5522.45    &  Fe~{\sc i}  &  4.209 &     $-$1.550   \\  
  5539.28    &  Fe~{\sc i}  &  3.642 &     $-$2.660   \\  
  5543.15    &  Fe~{\sc i}  &  3.695 &     $-$1.570   \\  
  5543.94    &  Fe~{\sc i}  &  4.217 &     $-$1.140   \\  
  5546.99    &  Fe~{\sc i}  &  4.217 &     $-$1.910   \\  
  5576.09    &  Fe~{\sc i}  &  3.430 &     $-$0.894  \\  
  5584.77    &  Fe~{\sc i}  &  3.573 &     $-$2.320   \\  
  5636.70    &  Fe~{\sc i}  &  3.640 &     $-$2.610   \\  
  5638.26    &  Fe~{\sc i}  &  4.220 &     $-$0.870   \\  
  5641.43    &  Fe~{\sc i}  &  4.256 &     $-$1.063  \\  
  5662.52    &  Fe~{\sc i}  &  4.178 &     $-$0.573  \\  
  5691.50    &  Fe~{\sc i}  &  4.301 &     $-$1.520   \\  
  5701.55    &  Fe~{\sc i}  &  2.559 &     $-$2.216  \\  
  5856.09    &  Fe~{\sc i}  &  4.294 &     $-$1.570   \\  
  5859.58    &  Fe~{\sc i}  &  4.549 &     $-$0.620   \\  
  5862.35    &  Fe~{\sc i}  &  4.549 &     $-$0.365  \\  
  5916.25    &  Fe~{\sc i}  &  2.453 &     $-$2.994  \\  
  5930.18    &  Fe~{\sc i}  &  4.652 &     $-$0.251  \\  
  5934.66    &  Fe~{\sc i}  &  3.928 &     $-$1.170   \\  
  5956.69    &  Fe~{\sc i}  &  0.859 &     $-$4.605  \\  
  5976.78    &  Fe~{\sc i}  &  3.943 &     $-$1.290   \\  
  5984.81    &  Fe~{\sc i}  &  4.733 &     $-$0.280   \\  
  5987.07    &  Fe~{\sc i}  &  4.795 &     $-$0.556  \\  
  6003.01    &  Fe~{\sc i}  &  3.881 &     $-$1.120   \\  
  6024.06    &  Fe~{\sc i}  &  4.548 &     $-$0.052  \\  
  6056.01    &  Fe~{\sc i}  &  4.733 &     $-$0.460   \\  
  6078.49    &  Fe~{\sc i}  &  4.795 &     $-$0.370   \\  
  6137.00    &  Fe~{\sc i}  &  2.198 &     $-$2.950   \\  
  6157.73    &  Fe~{\sc i}  &  4.076 &     $-$1.260   \\  
  6187.99    &  Fe~{\sc i}  &  3.943 &     $-$1.720   \\  
  6200.31    &  Fe~{\sc i}  &  2.608 &     $-$2.450   \\  
  6315.81    &  Fe~{\sc i}  &  4.076 &     $-$1.710   \\  
  6322.69    &  Fe~{\sc i}  &  2.588 &     $-$2.446  \\  
  6330.85    &  Fe~{\sc i}  &  4.733 &     $-$1.158  \\  
  6336.82    &  Fe~{\sc i}  &  3.686 &     $-$0.856  \\  
\hline
\end{tabular}
\label{tab:line_list}
\end{center}
\end{table}

\begin{table}
\begin{center}
  \contcaption{}
  \begin{tabular}{lllr}
  \hline
   $\lambda$ & Element & $\chi$ & $\log gf$ \\ 
   (\AA)     &         & (eV)   &           \\ 
 \hline
  6344.15    &  Fe~{\sc i}  &  2.433 &     $-$2.923  \\  
  6469.19    &  Fe~{\sc i}  &  4.835 &     $-$0.770   \\  
  6495.74    &  Fe~{\sc i}  &  4.835 &     $-$0.940   \\  
  6498.94    &  Fe~{\sc i}  &  0.958 &     $-$4.699  \\  
  6574.23    &  Fe~{\sc i}  &  0.990 &     $-$5.023  \\  
  6609.11    &  Fe~{\sc i}  &  2.559 &     $-$2.692  \\  
  6627.55    &  Fe~{\sc i}  &  4.548 &     $-$1.500    \\  
  6703.57    &  Fe~{\sc i}  &  2.758 &     $-$3.100    \\  
  6713.75    &  Fe~{\sc i}  &  4.790 &     $-$1.410   \\  
  6725.36    &  Fe~{\sc i}  &  4.100 &     $-$2.210   \\  
  6726.67    &  Fe~{\sc i}  &  4.610 &     $-$1.050   \\  
  6733.15    &  Fe~{\sc i}  &  4.640 &     $-$1.580   \\  
  6745.97    &  Fe~{\sc i}  &  4.070 &     $-$2.710   \\  
  6750.16    &  Fe~{\sc i}  &  2.420 &     $-$2.655  \\  
  6753.47    &  Fe~{\sc i}  &  4.560 &     $-$2.350   \\  
  6786.86    &  Fe~{\sc i}  &  4.190 &     $-$1.900    \\  
  6806.86    &  Fe~{\sc i}  &  2.730 &     $-$3.140   \\  
  6810.27    &  Fe~{\sc i}  &  4.610 &     $-$1.000    \\  
  6820.37    &  Fe~{\sc i}  &  4.640 &     $-$1.160   \\  
  6839.84    &  Fe~{\sc i}  &  2.560 &     $-$3.450   \\  
  6843.66    &  Fe~{\sc i}  &  4.550 &     $-$0.860   \\  
  6855.72    &  Fe~{\sc i}  &  4.610 &     $-$1.710   \\  
  6857.25    &  Fe~{\sc i}  &  4.070 &     $-$2.150   \\  
  6858.16    &  Fe~{\sc i}  &  4.610 &     $-$0.950   \\  
  6862.50    &  Fe~{\sc i}  &  4.560 &     $-$1.430   \\  
  6864.31    &  Fe~{\sc i}  &  4.560 &     $-$2.290   \\  
  6880.63    &  Fe~{\sc i}  &  4.150 &     $-$2.250   \\  
  6898.29    &  Fe~{\sc i}  &  4.220 &     $-$2.230   \\  
  6936.50    &  Fe~{\sc i}  &  4.610 &     $-$2.230   \\  
  6945.20    &  Fe~{\sc i}  &  2.420 &     $-$2.460   \\  
  6951.25    &  Fe~{\sc i}  &  4.560 &     $-$1.050   \\  
  6960.32    &  Fe~{\sc i}  &  4.590 &     $-$1.900   \\  
  6971.94    &  Fe~{\sc i}  &  3.020 &     $-$3.340   \\  
  6978.86    &  Fe~{\sc i}  &  2.480 &     $-$2.490   \\  
  6988.53    &  Fe~{\sc i}  &  2.400 &     $-$3.420   \\  
  7000.62    &  Fe~{\sc i}  &  4.140 &     $-$2.130   \\  
  7010.35    &  Fe~{\sc i}  &  4.590 &     $-$1.860   \\  
  7022.96    &  Fe~{\sc i}  &  4.190 &     $-$1.110   \\  
  7024.07    &  Fe~{\sc i}  &  4.070 &     $-$1.940   \\  
  7038.23    &  Fe~{\sc i}  &  4.220 &     $-$1.130   \\  
  7083.40    &  Fe~{\sc i}  &  4.910 &     $-$1.260   \\  
  7114.56    &  Fe~{\sc i}  &  2.690 &     $-$3.930   \\  
  7142.52    &  Fe~{\sc i}  &  4.950 &     $-$0.930   \\  
  7219.69    &  Fe~{\sc i}  &  4.070 &     $-$1.570   \\  
  7221.21    &  Fe~{\sc i}  &  4.560 &     $-$1.220   \\  
  7228.70    &  Fe~{\sc i}  &  2.760 &     $-$3.270   \\  
  7284.84    &  Fe~{\sc i}  &  4.140 &     $-$1.630   \\  
  7306.57    &  Fe~{\sc i}  &  4.180 &     $-$1.550   \\  
  7401.69    &  Fe~{\sc i}  &  4.190 &     $-$1.600    \\  
  7418.67    &  Fe~{\sc i}  &  4.140 &     $-$1.440   \\  
  7421.56    &  Fe~{\sc i}  &  4.640 &     $-$1.690   \\  
  7447.40    &  Fe~{\sc i}  &  4.950 &     $-$0.950   \\  
  7461.53    &  Fe~{\sc i}  &  2.560 &     $-$3.450   \\  
  7491.66    &  Fe~{\sc i}  &  4.300 &     $-$1.010   \\  
  7498.54    &  Fe~{\sc i}  &  4.140 &     $-$2.100    \\  
  7507.27    &  Fe~{\sc i}  &  4.410 &     $-$1.030   \\  
  7531.15    &  Fe~{\sc i}  &  4.370 &     $-$0.640   \\  
  7540.44    &  Fe~{\sc i}  &  2.730 &     $-$3.750   \\  
  7547.90    &  Fe~{\sc i}  &  5.100 &     $-$1.110   \\  
  7551.10    &  Fe~{\sc i}  &  5.080 &     $-$1.630   \\  
  7568.91    &  Fe~{\sc i}  &  4.280 &     $-$0.900    \\  
  7582.12    &  Fe~{\sc i}  &  4.950 &     $-$1.600    \\  
  7583.80    &  Fe~{\sc i}  &  3.020 &     $-$1.930   \\  
\hline
\end{tabular}
\label{tab:line_list}
\end{center}
\end{table}

\begin{table}
\begin{center}
  \contcaption{}
  \begin{tabular}{lllr}
  \hline
   $\lambda$ & Element & $\chi$ & $\log gf$ \\ 
   (\AA)     &         & (eV)   &           \\ 
 \hline
  7710.36    &  Fe~{\sc i}  &  4.220 &     $-$1.112  \\
  7745.52    &  Fe~{\sc i}  &  5.080 &     $-$1.140   \\
  7751.11    &  Fe~{\sc i}  &  4.990 &     $-$0.740   \\
  7807.91    &  Fe~{\sc i}  &  4.990 &     $-$0.510   \\
  7844.55    &  Fe~{\sc i}  &  4.830 &     $-$1.670   \\
  7912.87    &  Fe~{\sc i}  &  0.860 &     $-$4.850   \\
  7955.70    &  Fe~{\sc i}  &  5.030 &     $-$1.110   \\
  7959.15    &  Fe~{\sc i}  &  5.030 &     $-$1.180   \\ 
  5264.81    &  Fe~{\sc ii} &  3.230 &     $-$3.120   \\ 
  5325.55    &  Fe~{\sc ii} &  3.221 &     $-$3.222  \\ 
  5414.07    &  Fe~{\sc ii} &  3.221 &     $-$3.750   \\ 
  5425.26    &  Fe~{\sc ii} &  3.199 &     $-$3.372  \\ 
  5991.38    &  Fe~{\sc ii} &  3.153 &     $-$3.560   \\ 
  6084.11    &  Fe~{\sc ii} &  3.199 &     $-$3.780   \\ 
  6149.26    &  Fe~{\sc ii} &  3.889 &     $-$2.800    \\  
  6247.56    &  Fe~{\sc ii} &  3.892 &     $-$2.329  \\  
  6432.68    &  Fe~{\sc ii} &  2.891 &     $-$3.685  \\  
  6456.38    &  Fe~{\sc ii} &  3.903 &     $-$2.100    \\  
  6516.08    &  Fe~{\sc ii} &  2.891 &     $-$3.450   \\  
  4806.98    &  Ni~{\sc i}  &  3.679 &     $-$0.640   \\  
  4852.55    &  Ni~{\sc i}  &  3.542 &     $-$1.070   \\  
  4904.41    &  Ni~{\sc i}  &  3.542 &     $-$0.170   \\  
  4913.97    &  Ni~{\sc i}  &  3.743 &     $-$0.630   \\  
  4946.03    &  Ni~{\sc i}  &  3.796 &     $-$1.290   \\  
  5003.73    &  Ni~{\sc i}  &  1.676 &     $-$3.130   \\  
  5010.93    &  Ni~{\sc i}  &  3.635 &     $-$0.870   \\  
  5032.72    &  Ni~{\sc i}  &  3.898 &     $-$1.270   \\  
  5082.34    &  Ni~{\sc i}  &  3.658 &     $-$0.630   \\  
  5155.13    &  Ni~{\sc i}  &  3.898 &     $-$0.650   \\  
  5435.86    &  Ni~{\sc i}  &  1.986 &     $-$2.590   \\  
  5462.49    &  Ni~{\sc i}  &  3.847 &     $-$0.930   \\  
  5589.36    &  Ni~{\sc i}  &  3.898 &     $-$1.140   \\  
  5593.73    &  Ni~{\sc i}  &  3.898 &     $-$0.840   \\  
  5625.31    &  Ni~{\sc i}  &  4.089 &     $-$0.700    \\  
  5682.20    &  Ni~{\sc i}  &  4.105 &     $-$0.499  \\  
  6111.07    &  Ni~{\sc i}  &  4.088 &     $-$0.830   \\  
  6175.36    &  Ni~{\sc i}  &  4.089 &     $-$0.559  \\  
  6186.71    &  Ni~{\sc i}  &  4.105 &     $-$0.960   \\  
  6191.17    &  Ni~{\sc i}  &  1.676 &     $-$2.353  \\  
  6223.98    &  Ni~{\sc i}  &  4.105 &     $-$0.970   \\  
  6378.25    &  Ni~{\sc i}  &  4.154 &     $-$0.830   \\  
  6586.31    &  Ni~{\sc i}  &  1.951 &     $-$2.810   \\  
  6767.78    &  Ni~{\sc i}  &  1.830 &     $-$2.060   \\  
  6772.32    &  Ni~{\sc i}  &  3.660 &     $-$0.960   \\  
  6842.04    &  Ni~{\sc i}  &  3.660 &     $-$1.440   \\  
  7001.55    &  Ni~{\sc i}  &  1.930 &     $-$3.650   \\  
  7030.02    &  Ni~{\sc i}  &  3.540 &     $-$1.700    \\  
  7110.91    &  Ni~{\sc i}  &  1.930 &     $-$2.910   \\  
  7381.94    &  Ni~{\sc i}  &  5.360 &     $-$0.050   \\  
  7422.29    &  Ni~{\sc i}  &  3.630 &     $-$0.140   \\  
  7525.12    &  Ni~{\sc i}  &  3.630 &     $-$0.670   \\  
  7555.61    &  Ni~{\sc i}  &  3.850 &     $-$0.046  \\  
  7574.05    &  Ni~{\sc i}  &  3.830 &     $-$0.610   \\  
  7715.58    &  Ni~{\sc i}  &  3.700 &     $-$0.980   \\  
  7727.62    &  Ni~{\sc i}  &  3.680 &     $-$0.300    \\  
  7797.59    &  Ni~{\sc i}  &  3.300 &     $-$0.820   \\  
  7826.76    &  Ni~{\sc i}  &  3.700 &     $-$1.870   \\ 
  4810.53    &  Zn~{\sc i}  &  4.078 &     $-$0.170   \\ 
\hline
\end{tabular}
\label{tab:line_list}
\end{center}
\end{table}

\bsp

\label{lastpage}

\end{document}